\documentclass[letterpaper]{jpconf}
\usepackage{amsmath}
\usepackage{amsfonts}
\usepackage{graphicx}
\begin{document}

\newcommand\be{\begin{equation}}
\newcommand\ee{\end{equation}}
\newcommand\bea{\begin{eqnarray}}
\newcommand\eea{\end{eqnarray}}
\newcommand\bseq{\begin{subequations}} 
\newcommand\eseq{\end{subequations}}
\newcommand\bcas{\begin{cases}}
\newcommand\ecas{\end{cases}}
\newcommand{\p}{\partial}
\newcommand{\f}{\frac}

\title{Deformed spaces and loop cosmology}

\author{Marco Valerio Battisti}

\address{ICRA and Phys. Dept. University of Rome ``Sapienza'', P.le A. Moro 5 00185 Rome, Italy}

\ead{battisti@icra.it}

\begin{abstract}
The non-singular bouncing solution of loop quantum cosmology is reproduced by a deformed minisuperspace Heisenberg algebra. This algebra is a realization of the Snyder space, is almost unique and is related to the $\kappa$-Poincar\'e one. Since the sign of the deformation parameter it is not fixed, the Friedmann equation of braneworlds theory can also be obtained. Moreover, the sign is the only freedom in the picture and these frameworks are the only ones which can be reproduced by our deformed scheme. A generalized uncertainty principle for loop quantum cosmology is also proposed. 
\end{abstract}

\section*{Introduction}
In the last years a wide interest in the analysis of the non-commutative framework is increased (for a review see \cite{DN01}). In particular, this approach can be considered a plausible candidate for describing physics at the Planck scale \cite{DFR} and can be related to the intuitions of doubly special relativity (DSR) \cite{AMS} which arises as a semi-classical limit of quantum gravity (see \cite{RovSmo} and the references therein). On the other hand, a prerequisite of any quantum theory of gravity is to solve the space-time singularities predicted by the Einstein theory of general relativity. One of the most important example is the big-bang singularity appearing in the standard model of cosmology, which Friedmann dynamics is expected to be modified by quantum effects in the regime of small scale factor. Loop quantum cosmology (LQC) \cite{LQC} leads to a resolution of the singularity replacing the big-bang by a big-bounce as soon as the matter energy density reaches the Planck scale, i.e. in this extreme region (quantum) gravity behaves repulsively \cite{APS}. 

From this perspective it is natural to investigate the fate of the cosmological singularity in a deformed (minisuper)-space framework \cite{BMeAl}. Here, we consider generalized commutation relations which leave undeformed the translation group and preserve the rotational invariance, and apply this framework to the Friedmann-Robertson-Walker (FRW) Universes (for more details see \cite{Bat}). As a result the effective Friedmann equation of LQC \cite{SV} naturally arises from the deformed scheme. Interesting, since the deformed Heisenberg algebra is fixed up to a sign, also the braneworlds Friedmann equation \cite{Roy} is predicted by our model. In other words, this deformed phase space can be regarded, from a phenomenological point of view, as an effective framework which is able to describe the results obtained in both LQC and Randall-Sundrum braneworlds scenario. In this deformed scheme, the different predictions of such theories can be easily understood considering the opposite sign in the deformation term. It is worth noting that, a generalized uncertainty principle naturally arises in our framework and, in the braneworlds-like case, it resembles the one predicted by string theory. At the same level, a generalized uncertainty relation can also be proposed for LQC. 

The paper is organized as follows. Section 1 is devoted to discuss a non-commutative space and its commutations relations. In Section 2 this framework is applied to the isotropic cosmological models analyzing the modifications induced on the Friedmann equation. Finally, in Section 3, implications of the picture are analyzed. Concluding remarks follow.      

\section{Deformed Heisenberg algebra}
Let us start by considering a $n$-dimensional deformed (non-commutative) Euclidean space such that commutator between the coordinates has the non-trivial structure
\be\label{snyalg}
[\tilde x_i,\tilde x_j]=\mp i\alpha M_{ij},
\ee
where with $\tilde x$ we refer to the non-commutative coordinates and $\alpha>0$ is the deformation parameter with dimension of a length square. We then demand that the rotation generators $M_{ij}=x_ip_j-x_jp_i$ satisfy the ordinary $SO(n)$ algebra and that the translation group is not deformed, i.e. $[p_i,p_j]=0$. Is then natural to assume the commutators between $M_{ij}$ and $\tilde x_k$, as well as between $M_{ij}$ and $p_k$, as undeformed. This way we deal with the (Euclidean) Snyder space \cite{Sny}. 

We now consider a general rescaling of the non-commutative coordinates $\tilde x_i$ in terms of a momentum-dependent function $\varphi\left(\alpha p^2\right)$ as
\be\label{relalg}
\tilde x_i=x_i\varphi\left(\alpha p^2\right),
\ee
i.e. we consider a realization of the algebra (\ref{snyalg}) in terms of ordinary phase space variables \cite{Mel}. It is not difficult to show that the function $\varphi$ is uniquely fixed (up to a sign) by our natural assumptions and reads $\varphi\left(\alpha p^2\right)=\sqrt{1\pm\alpha p^2}$. This way the commutator between non-commutative coordinates and momenta is given by
\be\label{xp}
[\tilde x_i,p_j]=i\delta_{ij}\sqrt{1\pm\alpha p^2}
\ee
and the ordinary Heisenberg algebra is recovered as soon as $\alpha\rightarrow0$. This deformed algebra closes in sense that all the Jacobi identities are satisfied and can be related to the $\kappa$-Poincar\'e one \cite{Mag}. Interest in non-commutative (or deformed) phase space arises in order to mathematically describe DSR and, in particular, some of DSR models can be formulated as generalizations of Snyder model \cite{Kov}. As last point, it is worth noting that in the case of minus sign, the momentum is limited from above since $p\in I\equiv\left(-1/\sqrt\alpha, 1/\sqrt\alpha\right)$. Therefore, in this case, we have a truncation of the phase space at $p^2\geq\alpha$.

\section{Friedmann dynamics in the deformed framework}
The FRW cosmological models are characterized by imposing the isotropy on the Cauchy surfaces which fill the space-time manifold. Isotropy reduces the phase space of GR to be two dimensional with coordinates $(a,p_a)$, where the scale factor $a$ is the only degree of freedom of the system. It describe the expansion of the Universe and the standard model of cosmology is based on this models \cite{Kolb}. The dynamics of these Universes can be obtained from the extended Hamiltonian
\be\label{extham}
\mathcal H_E=\f{2\pi G}3N\f{p_a^2}a+\f3{8\pi G}Nak-Na^3\rho+\lambda\pi,
\ee
where $\lambda$ is a Lagrange multiplier and the parameter $k$ can be zero or $\pm1$ leading to the flat, open or closed Universe respectively. The term $\lambda\pi$ is introduced since $\pi$, the momentum conjugate to the lapse function $N$, vanishes. In the expression above $\rho=\rho(a)$ denotes a generic energy density we have introduced into the dynamics. 

Let us now consider the modifications induced on the dynamics of the FRW models by the deformed Heisenberg algebra discussed above \cite{Bat}.  In particular, we assume the symplectic structure of the minisuperspace as deformed and thus the Poisson bracket (for any two-dimensional phase space function) appears to be
\be\label{pb}
\{F,G\}_\alpha=\left(\f{\p F}{\p a}\f{\p G}{\p p_a}-\f{\p F}{\p p_a}\f{\p G}{\p a}\right)\sqrt{1\pm\alpha p_a^2}.
\ee
We deal with a one-dimensional mechanical system and thus the only non-trivial commutators is given by 
\be\label{funpb}
\{a,p_a\}=\sqrt{1\pm\alpha p_a^2}.
\ee
Since the Poisson bracket $\{N,\pi\}=1$ is not affected by the deformations induced by the $\alpha$ parameter on the system, the equations of motion $\dot N=\{N,\mathcal H_E\}=\lambda$ and $\dot\pi=\{\pi,\mathcal H_E\}=\mathcal H=0$ remain unchanged\footnote{As usually, $\mathcal H=0$ is the scalar constraint and is obtained by requiring the primary constraint $\pi=0$ will be satisfied at all times.}. On the other hand, the equations of motion of the scale factor $a$ and its conjugate momentum $p_a$ become modified in such an approach via the deformed symplectic geometry (\ref{pb}) and read
\bea\label{eqapgup}
\dot a&=&\{a,\mathcal H_E\}_\alpha=\f{4\pi G}3N\f{p_a}a\sqrt{1\pm\alpha p_a^2},\\\nonumber
\dot p_a&=&\{p_a,\mathcal H_E\}_\alpha=N\left(\f{2\pi G}3\f{p_a^2}{a^2}-\f3{8\pi G}k+3a^2\rho+a^3\f{d\rho}{da}\right)\sqrt{1\pm\alpha p_a^2}.
\eea 
The equation of motion for the Hubble rate $(\dot a/a)$ can be obtained solving the scalar constraint $\mathcal H=0$ with respect to $p_a$ and then considering the first equation of (\ref{eqapgup}). Explicitly it becomes
\be\label{deffri}
\left(\f{\dot a}a\right)^2=\left(\f{8\pi G}3\rho-\f k{a^2}\right)\left[1\pm\f{3\alpha}{2\pi G}a^2\left(a^2\rho-\f3{8\pi G}k\right)\right].
\ee
We refer to this equation as the {\it deformed Friedmann equation} as it entails the modification arising from the deformed Heisenberg algebra previously analyzed. Of course, for $\alpha\rightarrow0$ the ordinary Friedmann equation is recovered. Let us consider the flat FRW Universe, i.e. the $k=0$ model. In this case the deformed equation (\ref{deffri}) appears to be
\be\label{modfri}
\left(\f{\dot a}a\right)^2=\f{8\pi G}3\rho\left(1\pm\f\rho{\rho_c}\right),
\ee
where $\rho_c=(2\pi G/3\alpha)\rho_P$ is the critical energy density, and $\rho_P$ denotes the Planck one. When the limit $\alpha\rightarrow0$ is taken into account, the critical energy density diverges leading to the ordinary dynamics. It is worth stressing that we have assumed the existence of a fundamental minimal length. In fact, as widely accepted, one of the most peculiar consequences of all promising quantum gravity theories is the existence of a fundamental cut-off length, which should be related to the Planck one \cite{gar}. Therefore, although this minimal length appears differently in distinct contexts, it is reasonable that the scale factor (the energy density) has a minimum (maximum) at the Planck scale.

The modified Friedmann equations (\ref{modfri}) are known in literature. The one with the $(-)$ deformation term is the effective equation incorporating the LQC effects on the FRW dynamics \cite{SV}. It can be obtained using the geometric formulation of quantum mechanics \cite{Tav}. On the other hand, the string inspired Randall-Sundrum braneworlds scenario leads to a modified Friedmenn equation as in (\ref{modfri}) with the $(+)$ sign \cite{Roy}. The opposite sign of the $\rho^2$-term in such an equation, is the well-known key difference between the effective LQC and the Randall-Sundrum framework. In fact, the former approach leads to a non-singular bouncing cosmology while in the latter, because of the positive sign, the Hubble rate can not vanish and the Universe can not experiences a bounce (or more generally a turn-around) in the scale factor.

\section{Quantum mechanical implications of the deformed picture}
As we have seen, the deformed algebra (\ref{xp}) leads to effective dynamics of loop and braneworlds cosmologies in the minus or plus case respectively. Let us now investigate some implications of the deformed framework in the quantum theory. Firstly, we have to stress that the ($\pm$)-frameworks are, of course, physically different. More precisely, the deformed Hilbert spaces $\mathcal F_\pm$ underling the algebras (\ref{funpb}) can be written as \cite{Bat}
\be
\mathcal F_\pm=L^2\left(\mathbb R(I),dp_a/\sqrt{1\pm\alpha p_a^2}\right), 
\ee
for the $\pm\alpha$ deformation of the ordinary Heisenberg theory, respectively. It is worth noting that these Hilbert spaces are unitarily inequivalent each other and also with respect to the ordinary one $L^2(\mathbb R,dp_a)$. This is not surprising since the deformation of the canonical commutation relations can be viewed, from the realization (\ref{relalg}), as an algebra homomorphism which is a non-canonical transformation and thus it can not be implemented as an unitary transformation. New features are then introduced at both classical and quantum level. (For an application of the formalism to the harmonic oscillator problem see \cite{Bat}.) 

A peculiar feature which deserves to be analyzed is the uncertainty principle underlying the deformed symplectic structure (\ref{pb}). The generalized uncertainty relation can be immediately obtained from the commutator (\ref{funpb}) and reads
\be\label{uncrel}
\Delta a\Delta p_a\geq \f 1 2|\langle\left(1\pm\alpha p_a^2\right)^{1/2}\rangle|,
\ee
by which, using the basic proprieties $\langle p_a^{2n}\rangle\geq\langle p_a^2\rangle$ and $\langle p_a^2\rangle=(\Delta p_a)^2+\langle p_a\rangle^2$, the boundary of the allowed region begin
\be
\Delta a=\f12\left|\left(\f{1\pm\alpha\langle p_a\rangle^2}{(\Delta p_a)^2}\pm\alpha\right)^{1/2}\right|.
\ee
As we can easily see, for an infinite uncertainty in momentum (or better when the relation $\Delta p_a\gg(\Delta p_a)^\star\equiv\sqrt{(1\pm\alpha\langle p_a\rangle)/\alpha}$ holds), the uncertainty in the scale factor $a$ no longer vanishes but approaches the minimal value of $\Delta a_\text{min}=\sqrt\alpha/2$. It is interesting to note that, $\Delta a_\text{min}$ is the global minimum in the $(+)$-sector, while $\Delta a_\text{min}=0$ is allowed in the $(-)$-one. This appears as soon as the dispersion on its conjugate momentum $\Delta p_a$ reaches the critical value of $(\Delta p_a)^\star$.

Summarizing, in the $(+)$-sector a nonzero minimal uncertainty in the particle (Universe) position (the scale factor) appears. The resulting implications are quite profound. In fact, it is no longer possible to spatially localize a wave function with arbitrary precision and then no physical states which are position eigenstates exist at all since they were only formal ones \cite{Kem}. As we have seen, this framework leads to the same Friedamnn equation of the Randall-Sundrum braneworlds scenario and thus, it is not unexpected that the plus deformed uncertainty relation (\ref{uncrel}) contains, at the leading order in $\alpha$, the string theory result $\Delta x\sim(1/\Delta p+l_s^2\Delta p)$ \cite{String}, in which the string length $l_s$ can be identify with $\sqrt{\alpha/2}$.

On the other hand, to obtain the Friedmann equation found in the effective loop cosmology, the $(-)$-deformed Heisenberg algebra is required. In this case a vanishing uncertainty in position is allowed and therefore the position eigenstates are true physical states (in the same sense of the ordinary quantum mechanics). However, differently from the Heisenberg framework, an infinite uncertainty in momentum is no longer required and $\Delta x_\text{min}=0$ appears as soon as the finite value $(\Delta p)^\star\propto1/\sqrt\alpha$ is considered (we also remember that in this scheme there is a cut-off on the momentum, i.e. $|p|\leq1/\sqrt\alpha$). This way, considering the minus relation (\ref{uncrel}) at the first order in $\alpha$, a generalized uncertainty principle for LQC can be proposed to be $\Delta x\sim|1/\Delta p-l_L^2\Delta p|$, where $l_L=\sqrt{\alpha/2}$ can be regarded as the loop cut-off length scale. 

\section*{Conclusions}
The equations of motion of the FRW models obtained in LQC and in the braneworlds scenario can be reproduced by a deformed Heisenberg algebra. This algebra in the unique one which is consistent, in the sense of the Jacobi identities, with the assumptions that both the translation and rotation groups are undeformed and that the commutator between the non-commutative coordinates is as in (\ref{snyalg}). Notably, it is also related to the $\kappa$-Poincar\'e algebra and the only freedom in (\ref{relalg}) lies in the $\pm$ sign. The $(+)$-framework leads to the effective Friedmann dynamics of Randall-Sundrum braneworlds scenario, while the opposite one to that of LQC. From this perspective, the former framework is such that a vanishing uncertainty in position (the scale factor of the Universe) is not longer allowed. On the other hand, the $(-)$-scheme implies that the zero uncertainty in position appears for a finite uncertainty in the momentum, proportional to the natural cut-off of the framework. Summarizing, a non-commutative (deformed) picture which leads, at phenomenological level, to the prediction of more general theories can be formulated. The validity and applicability of this model to more complicated (and physically interesting) arenas will deserve future investigations.

\end{document}